\documentclass[ aip,jap, jmp,amsmath,amssymb, reprint,]{revtex4-1}
\usepackage{graphicx}
\usepackage{dcolumn}
\usepackage{bm}
\usepackage{natbib}

\begin{document}
\preprint{AIP/123-QED}
\title{Diffusion of degenerate minority carriers in a $p$-type semiconductor}
\date{\today}
\author{C. P. Weber}\email{cweber@scu.edu}
\author{Eric A. Kittlaus}
\affiliation{Department of Physics, Santa Clara University, 500 El Camino Real, Santa Clara, CA 95053-0315}

\begin{abstract}
We report ultrafast transient-grating experiments on heavily $p$-type InP at 15 K. Our measurement reveals the dynamics and diffusion of photoexcited electrons and holes as a function of their density $n$ in the range $2\times10^{16}$ to $6\times10^{17}$ cm$^{-3}$. After the first few picoseconds the grating decays primarily due to ambipolar diffusion. While at low density we observe a regime in which the ambipolar diffusion is electron-dominated and increases rapidly with $n$, at high $n$ it appears to saturate at 34 cm$^2$/s. We present a simple calculation that reproduces the main results of our measurements as well as of previously published measurements that had shown diffusion to be a flat or decreasing function of $n$. By accounting for effect of density on charge susceptibility we show that, in $p$-type semiconductors, the regime we observe of increasing ambipolar diffusion is unique to heavy doping and low temperature, where both the holes and electrons are degenerate; in this regime the electronic and ambipolar diffusion are nearly equal. The saturation is identified as a crossover to ambipolar diffusion dominated by the majority carriers, the holes. At short times the transient-grating signal rises gradually. This rise reveals cooling of hot electrons and, at high photocarrier density, allows us to measure ambipolar diffusion of 110 cm$^2$/s in the hot-carrier regime.\\
\\
The following article appeared in \textit{Journal of Applied Physics} and may be found at\\ \url{http://link.aip.org/link/?JAP/113/053711} \\ \\
\textit{Copyright 2013 American Institute of Physics. This article may be downloaded for personal use only. 
Any other use requires prior permission of the author and the American Institute of Physics.} 
\end{abstract}
\maketitle

\section{Introduction}

Charge-carrier diffusion coefficients in semiconductors have been studied for decades, owing to their importance as parameters in the operation of semiconductor devices;\protect{\cite{Lowney1991, Yang1978, Liou1994, Neamen2003, Jungo2004}} indeed, diffusion currents sometimes greatly exceed drift currents. The diffusion of minority carriers may differ from the same carriers' diffusion coefficient when they are majority carriers,\cite{Bennett1983} and so must be determined independently. The difference may be particularly important when the majority carriers are degenerate.

Many devices contain degenerately-doped regions, such as the base of a bipolar junction transistor. Under conditions of high injection both majority and minority carriers may be degenerate, as occurs in the active layer of a diode laser. The diffusion coefficient of degenerate minority carriers, however, has not been explored. 

In this work we measure the diffusion of minority electrons by measuring $D_a$, the coefficient of ambipolar diffusion.\protect{\cite{vanRoosbroeck1953}} When an excitation such as the absorption of light creates equal numbers of excess electrons and holes, each species briefly diffuses according to its own density gradient. As the positive and negative charges become spatially separated, the resulting electric field, the Dember field, causes the more mobile electrons to be held back by the less mobile holes. The resulting conjoined motion of the two species is ambipolar diffusion. Because the degree of charge separation is small, ambipolar diffusion is usually described under an approximation of local charge neutrality.

Due to the flexibility of optical measurements, the use of photoexcited carriers has emerged as a prominent method for measuring ambipolar diffusion.\protect{\cite{Akiyama1994, Cameron1996, Zhao2008, Zhao2009, Hu2011, Scajev2011, Paget2012}} However, the density of photoexcited carriers can strongly influence the measured diffusion coefficients.\protect{\cite{Akiyama1994, Scajev2011, Paget2012}} Experiments on $n$-type semiconductors give a  consistent result: $D_a$ is either an increasing or a flat function of photocarrier density. For instance, recent work on $n$-GaAs quantum wells has shown $D_a$ (or the related ``ambipolar spin diffusion'') to be increasing in lightly doped wells\protect{\cite{Chen2012}} and flat in nominally undoped wells.\protect{\cite{Zhao2009, Hu2011}} Bulk $p$-type semiconductors, however, have been less studied, and the results vary: Paget \textit{et al}.\protect{\cite{Paget2012}} recently observed ambipolar diffusion in GaAs that strongly decreases with excitation density, while Zhao\protect{\cite{Zhao2008}} saw no density dependence in silicon. Here we report measurements of heavily Zn-doped InP and find $D_a$ first to rapidly increase and then to level off.

Electron-hole scattering has been shown to influence the dependence of $D_a$ on photocarrier concentration,\protect{\cite{Scajev2011}} particularly in $p$-type quantum wells at low doping.\protect{\cite{Akiyama1994}} However, for bulk material we find that even neglecting density-dependent scattering rates, we can harmonize all three behaviors seen in our and others' data---increase, decrease, and flat---in a simple calculation following the approach of van Roosbroeck.\protect{\cite{vanRoosbroeck1953}} We assume local charge neutrality, but account for the influence of carrier density on the charge susceptibility. The regime of increasing $D_a$ that we observe is unique to heavily-doped samples at low temperature. In this regime $D_a$ nearly equals the electrons' diffusion coefficient.

Our ultrafast measurement allows us to separate several physical processes by their time-scales. The slower part of our signal reveals ambipolar diffusion. The faster part, lasting just a few picoseconds, reveals the effects of electron cooling and trapping. In this regime we are able to measure a very rapid ambipolar diffusion of holes and hot electrons. In keeping with our interpretation of the longer-time data, this rapid diffusion is due primarily to the decreased charge susceptibility of hot electrons.

\section{Methods}

To measure the diffusion and dynamics of photocarriers we use an ultrafast transient-grating method. A pair of ``pump'' laser pulses are simultaneously incident on the sample. The two pulses are non-collinear and interfere; their absorption excites photocarriers in a sinusoidal pattern with wavelength $\Lambda$ and wavevector $q=2\pi/\Lambda$. By locally modifying the index of refraction, the photocarriers create a ``grating'' off of which time-delayed probe pulses diffract. As photocarriers recombine and diffuse, the grating amplitude decays at a rate
\begin{equation}\frac{1}{\tau(q)}=Dq^2+\frac{1}{\tau_0}.
\label{DiffusionEq}
\end{equation}
Here $D$ is the diffusion coefficient, and $\tau_0$ is the lifetime of spatially uniform excitation. Measuring $\tau(q)$ at several $q$ determines $D$. We measure the diffracted probe amplitude in a reflection geometry, improve the efficiency by heterodyne detection,\protect{\cite{Vohringer1995}} and suppress noise by 95-Hz modulation of the grating phase and lock-in detection.\protect{\cite{Weber2005}}

The pump and probe pulses come from a mode-locked Ti:Sapphire laser with wavelength near 800 nm and repetition rate of 80 MHz. The pulses are focused on the sample to a spot of 145 $\mu$m diameter. As the laser's fluence is varied, the probe pulses are always a factor of 6 weaker than the pump pulses. At 800 nm InP has an absorption length of order 0.3 $\mu$m and reflectivity of 0.3 (Ref. \protect{\onlinecite{Aspnes1983}}), so at our highest fluence each pair of pump pulses photoexcites electrons and holes at a mean density of $n_{\text{ex}}\approx2\times10^{17}$ cm$^{-3}$.

Implicit in Eq. \ref{DiffusionEq} is the assumption that $n$, and therefore $D_a$, is independent of position. The pump pulses' interference excites densities from 0 in the grating's troughs to $2n_{\text{ex}}$ at the peaks. However, the recombination time in $p$-type InP is 33 ns (Ref. \protect{\onlinecite{Rosenwaks1992}}), while our interpulse spacing is 12 ns, so a steady-state population of photocarriers accumulates. Diffusion spreads these carriers uniformly, raising the troughs to about $2n_{\text{ex}}$ and the peaks to $4n_{\text{ex}}$, and making $D_a$ roughly position-independent.\protect{\cite{NonsinusoidalNote}} The mean value $3n_{\text{ex}}$ corresponds, at our highest fluence, to an electronic quasi-Fermi energy $E_{\text{eF}}/k_B\approx380$ K. The steady-state photocarrier population assists in cooling hot electrons to the lattice temperature, since carrier-carrier thermalization is more rapid than thermalization through phonons.

The pump pulses create a carrier population within about 0.3 $\mu$m of the sample's surface. As the carriers diffuse inward some leave the probed region, causing the grating's diffraction efficiency to decay. This decay process may contribute to $1/\tau_0$ but will not influence the measurement of $D$ because the inward diffusion is insensitive to the grating's in-plane wavevector $q$.

All data we report are on a $p$-type sample,\protect{\cite{ptypesamplegrower}} Zn-doped to a room-temperature carrier concentration of $4.5\times10^{18}$ cm$^{-3}$, which corresponds to $E_F/k_B\approx190$ K. It is oriented (100) and has room-temperature resistivity $\rho=2.65\times10^{-2}$ {$\Omega\,$cm} and Hall mobility $\mu_H=53$ cm$^2/${V$\,$s}. Except where noted all transient-grating data are at 15 K.  

\section{Results and Discussion}

\begin{figure}
\includegraphics[width=3.25 in]{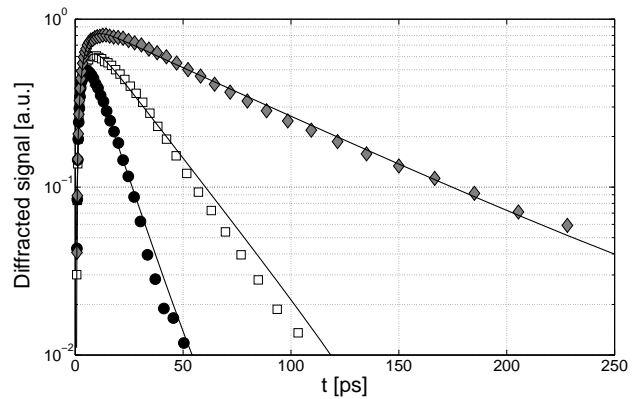}
\caption{Curves showing the time-dependence of the transient grating's amplitude for high-fluence (2.1 $\mu$J/cm$^2$) excitation. Values of $q$ are $1.26\times10^4$ cm$^{-1}$ (diamonds), $2.51\times10^4$ cm$^{-1}$ (squares), and $4.53\times10^4$ cm$^{-1}$ (circles). Lines are least-squares fits to Eq. \ref{RiseandFall} with $B=0$. Curves decay more quickly at high $q$, indicating carrier diffusion of 34 cm$^2$/s.}
\label{fallcurves}
\end{figure}

Figure \ref{fallcurves} shows typical time-dependence of the signal $S(t)$ diffracted from the transient grating after high-fluence excitation. $S(t)$ rises slowly, reaching its full amplitude only after several picoseconds, then decays exponentially. The time evolution of the transient-grating amplitude consists of three distinct pieces, well described by the equation

\begin{equation}S(t)=-Ae^{-t/\tau_{\text{rise}}}+Be^{-t/\tau_{\text{peak}}}+Ce^{-t/\tau_{\text{fall}}}.
\label{RiseandFall}
\end{equation}

Here $\tau_{\text{rise}}$ is the characteristic time of the curve's rising component, typically a few picoseconds; $1/\tau_{\text{fall}}$ is the rate of the grating's relaxation. The term $B\exp{(-t/\tau_{\text{peak}})}$ is zero except under very low-fluence excitation (Section \ref{peaksection}), in which case $\tau_{\text{peak}}$ has a value intermediate between $\tau_{\text{rise}}$ and $\tau_{\text{fall}}$. We begin in Section \ref{ambipolarsection} by discussing the dynamics of the falling component, and take up the rise and the peak in Section \ref{shortsection}.
 
\begin{figure}
\includegraphics[width=3.4 in]{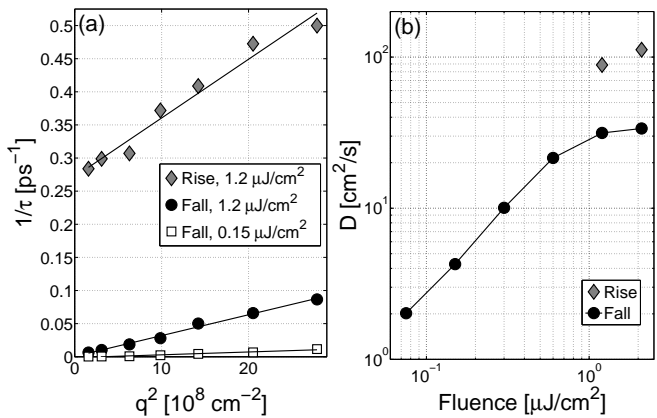}
\caption{\textbf{(a)} Grating decay rate $1/\tau$ \textit{vs}. $q^2$ at 15 K. Falling component: 1.2 $\mu$J/cm$^2$ (circles) and 0.15 $\mu$J/cm$^2$ (squares). Lines are fits to Eq. \ref{DiffusionEq}. The higher slope of the line at high fluence indicates significantly faster ambipolar diffusion than at low fluence. The rising component of the signal (diamonds, 1.2 $\mu$J/cm$^2$) diffuses still more quickly. \textbf{(b)} Diffusion as a function photoexcited density, as measured by excitation fluence. Diamonds: diffusion of the rising component at the two highest fluences. Circles: ambipolar diffusion of the falling component. Lines connect the points. At low fluence $D_a$ is electron-dominated and rises rapidly; at higher fluence an apparent saturation of $D_a$ is a crossover to hole-dominated diffusion.}
\label{DiffusionPlotsandDvsFlu}
\end{figure}

\subsection{Ambipolar Diffusion}\label{ambipolarsection}

By the time that the transient-grating signal begins to fall exponentially, the photoexcited electrons and holes have largely equilibrated to the lattice temperature. Electrons and holes are constrained to move together, resulting in ambipolar diffusion.

The lower two curves in Fig. \ref{DiffusionPlotsandDvsFlu}(a) show the decay rate $1/\tau_{\text{fall}}$ as a function of $q^2$. The agreement with Eq. \ref{DiffusionEq} reveals that carrier motion is diffusive. Strikingly, excitation at a fluence of 1.2 $\mu$J/cm$^2$ results in electrons and holes that diffuse much more quickly than they do when excited at 0.15 $\mu$J/cm$^2$. Figure \ref{DiffusionPlotsandDvsFlu}(b) shows the variation of $D_a$: at low fluence it rises swiftly with the density of photoexcited carriers, before saturating at high density. 

We explain the observed variation of $D_a$ in terms of the electron and hole mobilities $\mu_e$ and $\mu_h$, the background hole density $p_0$, and the density $n$ of photoexcited electrons and holes---which increases linearly with laser fluence. Under the assumption of local charge neutrality, one obtains the well-known result\protect{\cite{vanRoosbroeck1953}} for the ambipolar diffusion coefficient:

\begin{equation}D_a=\frac{\sigma_eD_h+\sigma_hD_e}{\sigma_e+\sigma_h}.
\label{standardambipolar}
\end{equation}
$D_e$ and $D_h$ are the electron and hole diffusion coefficients; $\sigma_e$ and $\sigma_h$ are the conductivities due to only the electrons or holes, respectively. 

Let $\mu_{\text{chem}}^e$ be the electrons' chemical potential, and define the electronic susceptibility as $\chi_e\equiv dn/d\mu_{\text{chem}}^e$; define hole susceptibility $\chi_h$ analogously. Then diffusion is related to mobility by:\protect{\cite{Finkelshtein1983, Castellani1987}}

\begin{equation}D_{e}=\frac{\sigma_{e}}{e^2\chi_{e}}=\frac{n\mu_{e}}{e\chi_{e}},
\label{EinsteinRelation}
\end{equation}
with an analogous equation for holes. These equations are just the Einstein relation, written in a form that is valid at all temperatures. In three dimensions $\chi_e$ must be evaluated numerically,\protect{\cite{Mohankumar1995}} but in the non-degenerate limit Eq. \ref{EinsteinRelation} reduces to the familiar expression ${D_e=\mu_ek_BT/e}$. In the degenerate limit 
\begin{equation}D_e=\frac{n\mu_{e}}{eN(E_{\text{eF}})}\propto n^{2/3}\mu_e, 
\end{equation}
where $N$ is the density of states. 

Combining Eqs. \ref{standardambipolar} and \ref{EinsteinRelation},

\begin{equation}D_a=\frac{(p_0+n)\mu_e\mu_h}{e\mu_e+e\mu_h\left(1+\frac{p_0}{n}\right)}\left(\frac{1}{\chi_e}+\frac{1}{\chi_h}\right).
\label{AmbipolarVsFlu}
\end{equation}
We can write the last factor as $1/\chi^*\equiv1/\chi_e+1/\chi_h$. As $n$ increases with fluence, $\mu_e$ will decrease (as will $\mu_h$, more slowly); but this decrease is typically quite weak.\protect{\cite{Ioffe}} If we treat $\mu_{e,h}$ as constants, Eq. \ref{AmbipolarVsFlu} can explain our observation that $D_a$ rises and then saturates. In fact $D_a$ will vary non-monotonically, giving varied behaviors that match the variety of trends observed in $D_a$ \textit{vs.} fluence.\protect{\cite{Chen2012, Hu2011, Paget2012, Zhao2008, Zhao2009}} 

To gain insight into Eq. \ref{AmbipolarVsFlu}, we evaluate it first in several limiting cases, then numerically. When ${n\ll p_0}$, at low fluence, ${D_a=n\mu_e/e\chi^*}$. This result differs from the electrons' unipolar diffusion coefficient $D_e$ by the factor ${\chi_e/\chi^*}$, but at low fluence, $\chi_e\approx\chi^*$. For sufficiently high $n$ electronic degeneracy causes $D_a$ to increase with $n$, as observed. To reach high $n$ while maintaining $n\ll p_0$, as at our lowest fluences, requires a heavily-doped sample.

At fluences much higher than those used in this experiment, ${n\gg p_0}$, the diffusion is proportional to the hole mobility and increases with $n$: ${D_a=n\mu_h/e\chi^*}$. The crossover from electron-dominated to hole-dominated ambipolar diffusion results in the apparent saturation of $D_a$ seen in our data [Fig. \ref{DiffusionPlotsandDvsFlu}(b)] and calculations [Fig. \ref{Prediction}(b)]. If the ratio $b\equiv\mu_e/\mu_h$ is sufficiently large, the crossover region may be approximated by both ${n\ll p_0}$ and ${n\mu_e\gg p_0\mu_h}$. In this case ${D_a=p_0\mu_h/e\chi^*}$, giving $D_a\propto n^{-1/3}$ at low temperature. In fact for InP $b$ is of order 10 (Ref. \protect{\onlinecite{Ioffe}}), so the crossover is narrow enough to appear flat rather than decreasing. 

\subsubsection{Numerical evaluation of Eq. \ref{AmbipolarVsFlu}}

\begin{table*} 
\begin{ruledtabular}
\begin{tabular}{ll|ccccc|l|cc}
Reference & Material & $p_0$ [cm$^{-3}$] & $T$ [K] & $m^*_e$ & $m^*_h$ & $\mu_e/\mu_h$ & $D_a$ \textit{vs}. $n$ measured & Fig.  \ref{80KandLiterature}(b) line & Data Fig. \\ \hline \hline
Present work & Zn:InP & $4.5\times10^{18}$ & 15 & .079 & .6 & 10 & Increase & 1 & \ref{DiffusionPlotsandDvsFlu}(b) \\
 &  &  & 80 &  &  &  & Increase (slight) & 2 & \ref{80KandLiterature}(a) \\ \hline
\protect{\onlinecite{Zhao2008}} Zhao & B:Si & $10^{15}$ & 90 & .36 & .81 & 3 & Flat & 3 & \\
 &  &  & 300 &  &  &  & Flat & 4 & \\ \hline
 \protect{\onlinecite{Paget2012}} Paget \textit{et al.} & Be:GaAs & $10^{17}$ & 300 & .063 & .53 & 10 & Decrease & 5 & \\
\end{tabular}
\end{ruledtabular}\caption{Parameters used in calculating $D_a/\mu_h$, taken from Refs. \protect{\cite{Paget2012, Zhao2008, Ioffe, Lowney1991}}. $m^*_e$ and $m^*_h$ are density-of-states effective masses in units of the free electron mass. Calculations of $\chi$ assume parabolic, isotropic bands.}
\label{tableparams}
\end{table*}

\begin{figure}
\includegraphics[width=3.25 in, height=2.5 in]{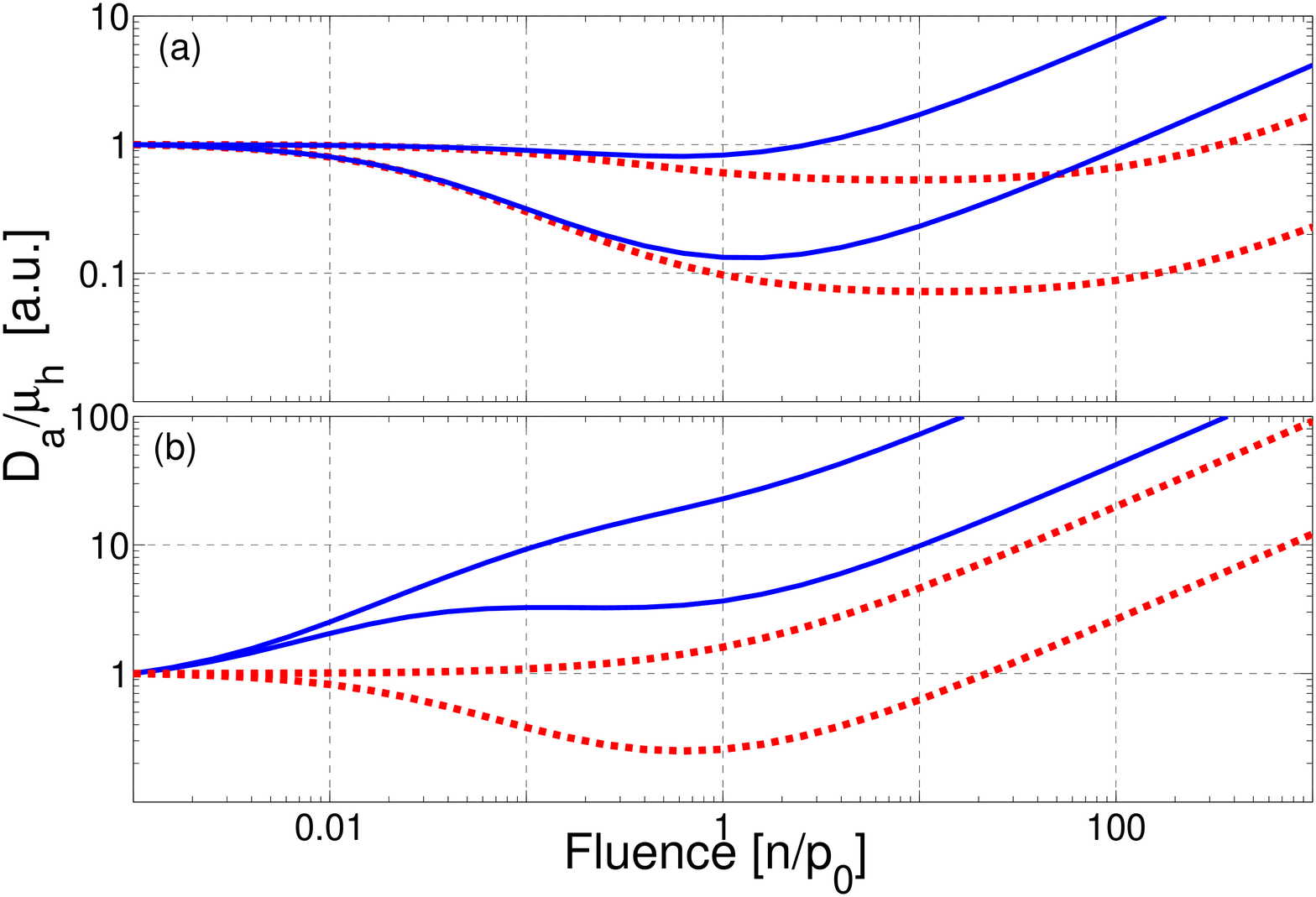}
\caption{(Dual-log scales; color online) Numerical calculations for $D_a/\mu_h$, for $T=$ 15 K (solid) and 300 K (dashed). For each panel, curves are $b=3$ (higher) and 30 (lower). Each curve is normalized to its low-fluence value. \textbf{(a)} Lightly doped, $p_0=10^{16}$ cm$^{-3}$. \textbf{(b)} $p_0=4.5\times10^{18}$ cm$^{-3}$, corresponding to the sample measured in this work.}
\label{Prediction}
\end{figure}

To further elucidate our and others' results on $p$-type semiconductors, we evaluate Eq. \ref{AmbipolarVsFlu} numerically. To avoid the complication of density-dependent mobilities, we plot the ratio $D_a/\mu_h$ for a few representative values of $b$, using the InP electron and hole masses shown in Table \ref{tableparams}. The results for lightly-doped $p$-type material at 15 K and 300 K appear in Fig. \ref{Prediction}(a). For both temperatures, at low photocarrier density $D_a/\mu_h=bk_BT/e$ is a constant. It decreases in the crossover region, then increases in the hole-dominated, high-fluence regime. (By contrast, Eq. \ref{AmbipolarVsFlu} shows that for $n$-type samples $D_a$ never decreases with $n$, consistent with measurements in nominally undoped\protect{\cite{Zhao2009, Hu2011}} and Si-doped\protect{\cite{Chen2012}} GaAs quantum wells.)

Figure \ref{Prediction}(b) shows $D_a/\mu_h$ calculated for a sample doped as heavily as ours. For low temperature, even at low excitation density the degeneracy of photoexcited electrons causes $D_a/\mu_h$ to increase with $n$; this behavior does not occur at high temperature, or in lightly-doped samples at any temperature. The crossover regime manifests as the saturation seen in our highest-fluence data. Beyond this saturation, in the hole-dominated regime $D_a/\mu_h$ continues to increase. 

\begin{figure}
\includegraphics[width=3.375 in]{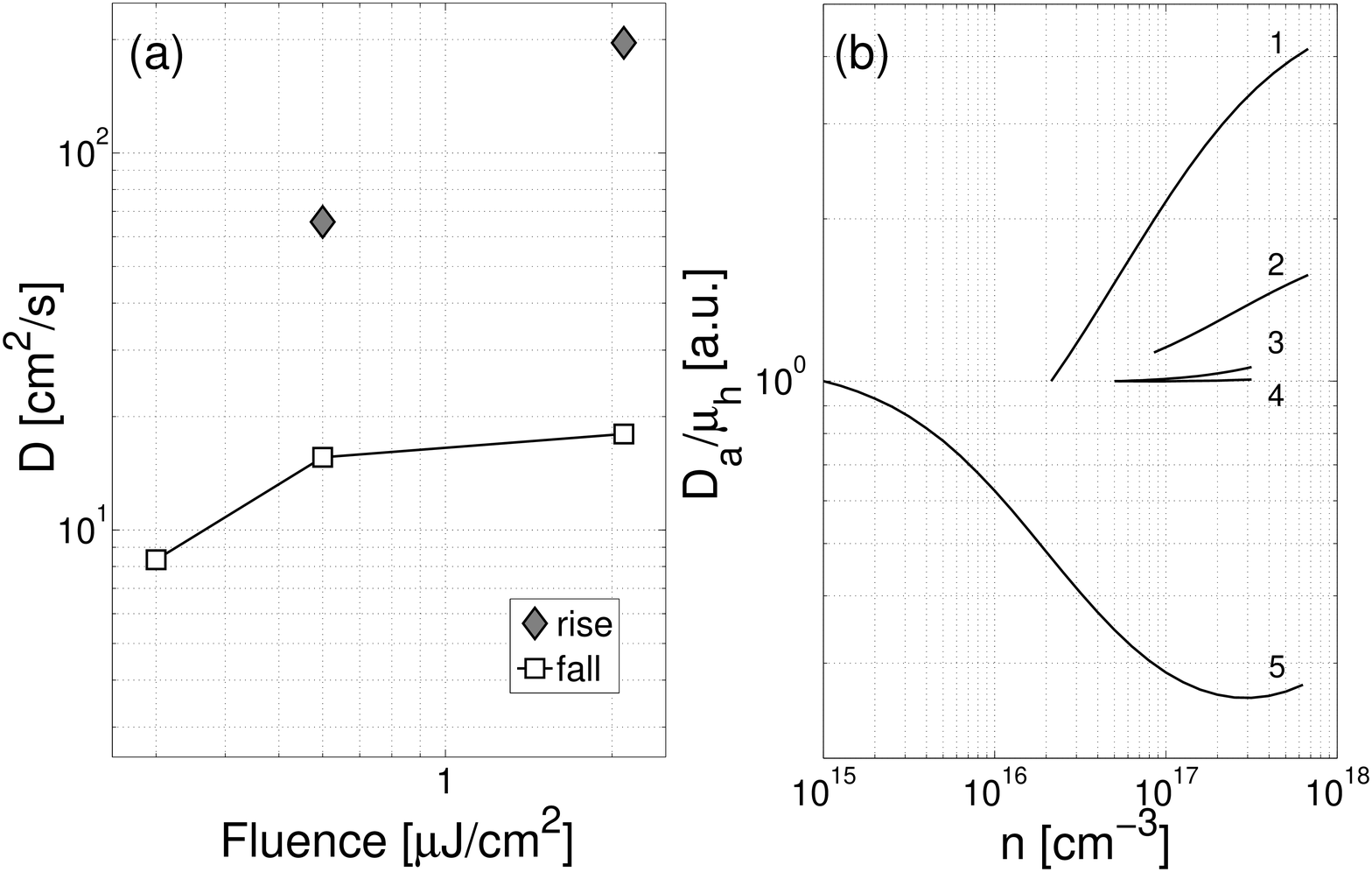}
\caption{\textbf{(a)} Measured diffusion \textit{vs}. fluence at 80 K. $D_a$, as expected, increases more slowly than at 15 K. The diffusion of the rise remains much faster than that of the fall. \textbf{(b)} $D_a/\mu_h$ normalized to its low-fluence value, calculated using parameters from Table \ref{tableparams}. Under the conditions of: our experiment at 15 K (1) and 80 K (2); Zhao at 90 K (3) and 300 K (4); and Paget \textit{et al}. (5). Calculated trends agree with those observed.}
\label{80KandLiterature}
\end{figure}

For heavy doping and 300 K [Fig. \ref{Prediction}(b), dashed lines], the calculated $D_a/\mu_h$  does not increase at low fluence. Low signal precludes measuring our sample at 300 K, but we have measured $D_a$ at 80 K [Fig. \ref{80KandLiterature}(a)]. It confirms our expectation by behaving intermediately between the high- and low-temperature predictions, increasing more weakly than at 15 K. 

Turning to others' data, Fig. \ref{80KandLiterature}(b) shows the evaluation of Eq. \ref{AmbipolarVsFlu} under the conditions of Refs. \protect{\onlinecite{Zhao2008}} and \protect{\onlinecite{Paget2012}}. With increasing fluence, these papers found $D_a$ to be unchanging and strongly decreasing, respectively. Accounting for the experiments' doping levels and material parameters (Table \ref{tableparams}), the calculations reproduce both trends. The figure also shows calculated curves for the specific conditions of our own measurements, using our best estimate of the mobility ratio, $b=10$. The 80 K prediction for $D_a$ does increase weakly with $n$, as observed.

\subsection{Short-time dynamics}\label{shortsection}
\subsubsection{Rise}

\begin{figure}
\includegraphics[width=3 in]{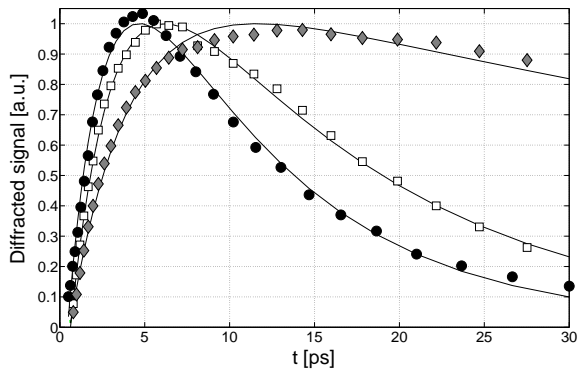}
\caption{High-fluence (2.1 $\mu$J/cm$^2$) transient-grating curves. Values of $q$ are $1.26\times10^4$ cm$^{-1}$ (diamonds), $3.77\times10^4$ cm$^{-1}$ (squares), and $5.28\times10^4$ cm$^{-1}$ (circles). Lines are least-squares fits to Eq. \ref{RiseandFall} with $B=0$. Higher-$q$ curves rise more quickly, indicating that carrier motion is diffusive even at short times. The diffusion coefficient of the rising component, 110 cm$^2$/s, is much faster than the diffusion of the falling component.}
\label{risecurves}
\end{figure}

The transient grating's behavior at short times, while electrons are still hot, further supports the analysis presented above. Figure \ref{risecurves} shows the signal for high-fluence excitation. At higher $q$ the curve not only falls more quickly, but also rises more quickly. A plot of $1/\tau_{\text{rise}}$ \textit{vs}. $q^2$ [upper curve of Fig. \ref{DiffusionPlotsandDvsFlu}(a)] reveals diffusion that is much faster (3.2 times greater) than the ambipolar diffusion of the fall. 

The increased ambipolar diffusivity of hot carriers arises from two factors. First, in polar, cubic semiconductors hot electrons' diffusivity has been observed to exceed that of cold electrons by factors up to five.\protect{\cite{Ruch1968}} Under optical excitation holes are heated much less than electrons, so we suppose that $\mu_e$ increases while $\mu_h$ remains unchanged; then Eq. \ref{AmbipolarVsFlu} shows that electronic mobility can increase $D_a$ by a factor no greater than 1.6, because in the crossover regime at high fluence $D_a$ is strongly influenced by hole mobility. The remaining increase comes from $\chi_e$, which depends on the electrons' density and temperature. A modest electronic heating of about 45 meV, for instance, would make the electrons non-degenerate and increase $1/\chi^*$ by a factor of two.

The electrons' initially high temperature may also give the rise its distinctive shape: a sign opposite to the rest of $S(t)$, and a short lifetime. Our pump and probe photons have the same energy, and the diffracted signal $S(t)$ arises from transient, local changes to the complex index of refraction evaluated at the probe's energy. The pump pulse initially excites electrons to states above the conduction-band minimum. As the electrons cool they move to lower energy states that are no longer degenerate with the probe-photon energy. We speculate that this cooling may be accompanied by a shift in the primary origin of $S(t)$---for instance from phase-space filling to bandgap renormalization or free-carrier absorption. The latter two mechanisms are expected to yield a sign opposite to that of phase-space filling.\protect{\cite{Kumar2011}

The rise's lifetime of a few picoseconds reflects the electrons' rapid thermalization. This thermalization slows at high fluence [Fig. \ref{RiseandPeak}(a)], in keeping with many,\protect{\cite{Zhou1989,Hohenester1993,Carmody2002}} but not all,\protect{\cite{Tsai2007}} prior observations. The slowing is attributed to the hot-phonon effect:\protect{\cite{Potz1983}} hot electrons couple primarily to LO phonons. The LO phonons couple weakly to acoustic phonons, giving heat only slowly to the lattice but frequently back to the electrons. This effect is particularly strong in InP because of the wide energy gap between acoustic and optical phonon branches.\protect{\cite{Clady2012}} The slowing may be compounded by increased screening\protect{\cite{DasSarma1987,Wen2006}} or by laser heating, as seen by comparing the rise lifetime at 15 K with that at  80 K. The rise at 80 K [Figs. \ref{80KandLiterature}(a),  \ref{RiseandPeak}(a)], though longer-lived, behaves the same as at low temperature: at high fluence it diffuses much more quickly than does the fall. As at low temperature, the increased diffusivity owes more to the hot electrons' susceptibility than to their scattering rate.

\subsubsection{Peak}\label{peaksection}

\begin{figure}
\includegraphics[width=3.35 in]{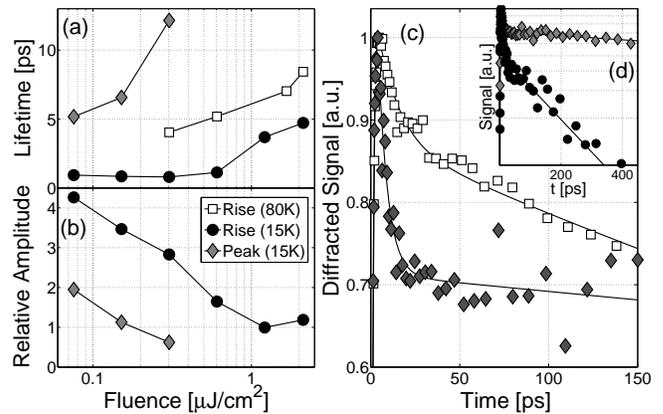}
\caption{\textbf{(a)} As a function of excitation fluence, lifetime of the rising component at 15 K (circles), rising component at 80 K (squares), and peak at 15 K (diamonds). Both features decay more slowly at higher fluence. Lines connect the points. \textbf{(b)} Ratio at 15 K of the coefficients $A/C$ (circles) and $B/C$ (diamonds) in Eq. \ref{RiseandFall}. At higher fluence the rise and peak features are both suppressed relative to the falling component. Ratios are the mean over all measured $q$. \textbf{(c)} Low-fluence transient-grating signal (suppressed zero) at 15 K. Squares: 0.3 $\mu$J/cm$^2$, $q=1.76\times10^4$ cm$^{-1}$. Diamonds: 0.08 $\mu$J/cm$^2$, $q=1.26\times10^4$ cm$^{-1}$. In the lower-fluence data a prominent peak-like feature precedes the slow decay. At slightly higher fluence the peak begins to blend with the slow decay. Lines are least-squares fits to Eq. \ref{RiseandFall} with $B\neq0$.  \textbf{(d)} 15 K transient-grating curves (log scale) at  0.08 $\mu$J/cm$^2$. Diamonds: $q=1.26\times10^4$ cm$^{-1}$. Circles: $q=4.53\times10^4$ cm$^{-1}$. Despite the complicating presence of the peak, the diffusion of the fall is clearly visible, though only 2 cm$^2$/s.}
\label{RiseandPeak}
\end{figure}

In addition to the change in diffusion rate $D_a$ with excitation fluence, the transient-grating signal's shape changes. Most markedly, at fluences of of 0.3 $\mu$J/cm$^2$ and lower, $S(t)$ acquires an additional component at short times, which appears as a ``peak,'' shown in Figs. \ref{RiseandPeak}(c) and \ref{RiseandPeak}(d). Fits to the form of Eq. \ref{RiseandFall} then require $B\neq0$. 

As fluence increases, the size of the peak decreases relative to the overall signal size [Fig. \ref{RiseandPeak}(b)]. This is consistent with the peak arising from the trapping of a fixed number (and therefore a decreasing proportion) of carriers. Because the peak appears at low fluence, where signal is small, we cannot determine whether it diffuses, but we plot its lifetime (the mean over all $q$ measured) in Fig. \ref{RiseandPeak}(a). The peak's lifetime increases with increasing fluence. This slowing may arise from the electrons' slower thermalization: the cross-section for trapping of high-energy electrons is lower than that for low-energy ones.\protect{\cite{MarionThornton}} We believe the peak does not disappear abruptly at high fluence, but its reduced size and its slower decay renders it indistinguishable from the main decay process, $\tau_{\text{fall}}$, due to ambipolar diffusion. 

Most important to the present work, the existence of the peak does not introduce any ambiguity into our low-fluence measurements of $D_a$. The ambipolar diffusion takes place during the signal's slow fall. As Fig. \ref{RiseandPeak}(d) shows, after the peak has completely decayed the slow fall is the signal's only remaining component.

\section{Conclusions}

We performed transient-grating measurements of ambipolar diffusion in $p$-type InP for a wide range of photoexcited carrier densities. Ambipolar diffusion is commonly described by two ``rules of thumb'': that $D_a$ is controlled by the minority carriers,\protect{\cite{Yang1978, Liou1994, Neamen2003}} or by the less-mobile carriers;\protect{\cite{Smith1989, Poortmans2006, Li2011}} in a $p$-type sample both cannot simultaneously be true. In fact when $n\ll p_0$, at low fluence,  $D_a$ is always nearly $D_e$. Our measurements in this regime revealed diffusion that increases strongly with $n$ due to the electrons' degeneracy. 

As $n$ increases our measurements reach a regime of  intermediate density where $D_a$ is proportional to hole mobility but strongly enhanced by the factor $\chi_h/\chi_e$. Here the ambipolar diffusion levels off, but the several-picosecond rise of the transient grating's amplitude reveals the distinct and much faster ambipolar diffusion of holes and hot electrons. The rise diffusion reaches 110 cm$^2$/s, due largely to the effect of heating on the electrons' susceptibility $\chi_e$.

We presented a simple calculation that assumes local charge neutrality and neglects the density-dependence of carrier mobilities. This calculation reproduces the salient features of our data---$D_a$'s initial rise and subsequent leveling off---provided one accounts for the electrons' degeneracy through $\chi_e$. It also reproduces the $D_a$ seen by Zhao\protect{\cite{Zhao2008}} and Paget \textit{et al}.\protect{\cite{Paget2012}} in more lightly-doped $p$-type samples, which are level and falling, respectively, over a broad range of excited densities.

At yet higher electronic density Eq. \ref{AmbipolarVsFlu} (Fig. \ref{Prediction}) predicts that $D_a$ will increase with $n$ for any temperature or doping level. At such high fluence $D_a$ may be dominated by the less-mobile carriers,\protect{\cite{HighFluNote}} but this regime in $p$-type samples has yet to be observed.

Previous optical measurements have not seen diffusion of degenerate minority carriers, even though---as is evident from Fig.  \ref{80KandLiterature}(b)---optical experiments routinely excite electronic densities corresponding to $E_{\text{eF}}/k_B\geq300$ K, and experiments with amplified lasers reach densities far higher. Rather, electronic degeneracy is usually obscured by the crossover to hole-dominated diffusion that begins when $n\mu_e$ is of order $p_0\mu_h$. Diffusion coefficients under conditions of heavy doping and high injection are important to the operation of many semiconducting devices. Our transient-grating measurement highlights the important role of degeneracy in determining diffusivity, even for minority carriers.

\begin{acknowledgements}
This work was supported by the National Science Foundation Grant No. DMR-1105553. E.A.K. was partly supported by Santa Clara University's Hayes Scholarship.
\end{acknowledgements}


\begin{thebibliography}{38}
\bibitem {NonsinusoidalNote}%
  \BibitemOpen
  \href@noop {} {}\bibinfo {note} {{T}he diffusion is indeed larger at the peaks than in the troughs, but numerical simulations confirmed that this effect modifies the grating's decay rate by no more than 4\%.}\BibitemShut {Stop}%
\end{thebibliography}
\end{document}